\documentclass[epj]{webofc}
\pdfoutput=1
\usepackage[utf8]{inputenc}
\usepackage[varg]{txfonts}   
\usepackage{booktabs}
\usepackage{xcolor}
\definecolor{darkred}{rgb}{0.4,0.0,0.0}
\definecolor{darkgreen}{rgb}{0.0,0.4,0.0}
\definecolor{darkblue}{rgb}{0.0,0.0,0.4}
\usepackage[bookmarks,linktocpage,colorlinks,
    linkcolor = darkred,
    urlcolor  = darkblue,
    citecolor = darkgreen]{hyperref}
\usepackage{tikz}
\usetikzlibrary{decorations.pathreplacing}
 \usetikzlibrary{plotmarks}
\usepackage{subfigure}
\usepackage{graphicx}
\wocname{EPJ Web of Conferences}
\woctitle{Lattice2017}
\begin{document}
%
\selectlanguage{english}
\title{%
Distribution of the Dirac modes in QCD
}
\author{%
\firstname{M.} \lastname{Catillo}\inst{1}
\and
\firstname{L. Ya.} \lastname{Glozman}\inst{1} 
}
\institute{%
Universität Graz, Institut für Physik, Universitätsplatz 5, 8010 Graz, Austria
}
\abstract{%
 It was established that distribution of the near-zero
 modes of the Dirac operator is consistent with the 
 Chiral Random Matrix Theory (\emph{CRMT}) and can be considered as
 a consequence of spontaneous breaking of chiral symmetry (\emph{SBCS})
 in QCD. The higher-lying modes of the Dirac operator carry information
 about confinement physics and are not affected by \emph{SBCS}. We study
 distributions of the near-zero and higher-lying modes of the
 overlap Dirac operator within $N_F = 2$ dynamical simulations. We
 find that distributions of both near-zero and higher-lying modes
 are the same and follow the Gaussian Unitary Ensemble of Random
 Matrix Theory. This means that randomness, while consistent with
 \emph{SBCS}, is not a consequence of \emph{SBCS} and is related to some more
 general property of QCD in confinement regime. 
}
\maketitle
\section{Introduction}\label{intro}

The near-zero modes of the Dirac operator in QCD are related to spontaneous
breaking of chiral symmetry (\emph{SBCS}) via the Banks-Casher relation 
\cite{BANKS1980103}. On the lattice with a finite volume they are
subject to some universal behaviour \cite{PhysRevD.46.5607}.
Within the $\epsilon$-regime, i.e. when $\Lambda_{QCD}L \gg 1$ and $m_{\pi}L \ll 1$, where $L$ is the linear size of the lattice and $m_{\pi}$ is the  pion mass, the Chiral Random Matrix Theory (\emph{CRMT}) links the distribution law
of the near-zero modes of the Dirac operator with the random matrices
\cite{SHURYAK1993306,537280920001201}. 

With $N_F$
degenerate quark flavors of mass $m$ the Dirac operator in the Weyl
representation of $\gamma$-matrices can be written as

\begin{equation}
D = \left( \begin{matrix} m & iW \\ iW^{+} & m   \end{matrix}\right).
\end{equation}

\noindent
The distribution of the large matrix $W$ with dimension determined by the
lattice size,  is given by 

\begin{equation}
 P(W) = \mathcal{N}  det(D)^{N_f} e^{-\frac{N\beta \Sigma^2}{4}tr(W^{\dagger}W)},
\label{eq-P(W)}
\end{equation}

\noindent
where $\mathcal{N}$ is the normalization constant, $\Sigma$ is a parameter that it is not always related to the chiral condensate (not - if we are beyond the $\epsilon$ regime),
and $\beta$ is the Dyson index which is determined by the symmetry properties of the matrix $W$. 
For different values of $\beta$ we have different matrix ensembles.
If $\beta =1$ we have the \emph{ Gaussian Orthogonal Ensemble} (\emph{GOE}), if $\beta =2$ the \emph{ Gaussian Unitary Ensemble} (\emph{GUE}) and 
$\beta = 4$, the \emph{ Gaussian Symplectic Ensemble} (\emph{GSE}). 
In \emph{QCD} $\beta=2$ as was shown in Ref. \cite{PhysRevLett.72.2531}.

Given eq. (\ref{eq-P(W)}) properties of the near-zero modes in QCD
are connected with properties of large random matrices. The near-zero
mode distribution has been studied on the lattice in many papers,
see e.g. \cite{PhysRevD.76.054503}, and a perfect agreement with
the \emph{CRMT} has been found.

While the lowest-lying modes of the Dirac operator are strongly affected by
\emph{SBCS}, the higher-lying modes are subject to confinement physics. This was
recently observed on the lattice via truncation of the lowest modes of the
overlap Dirac operator from quark propagators \cite{PhysRevD.89.077502,PhysRevD.91.034505,PhysRevD.91.114512,
PhysRevD.92.099902}. Hadrons  survive this truncation (except for pion)
and their mass remains large. It was noticed that not only
 $SU(2)_L \times SU(2)_R$ and $U(1)_A$ chiral symmetries get restored,
 but actually some higher symmetry appears. This symmetry was established
 to be $SU(2)_{CS}$ (chiral-spin) and $SU(4)$ that contains chiral
 symmetries as subgroups and that is a symmetry of confining
 chromo-electric interaction \cite{Glozman2015,PhysRevD.92.016001}. Given  lattice
 size $L \sim 2 $ fm in refs. \cite{PhysRevD.89.077502,PhysRevD.91.034505,PhysRevD.91.114512,
PhysRevD.92.099902} it was enough to remove lowest 10-20 modes
of the Dirac operator to observe emergence of 
$SU(2)_{CS}$  and $SU(4)$. Consequently the higher-lying modes,
that are above the lowest 10-20 modes, are not affected by \emph{SBCS} and by
breaking of $U(1)_A$ 
and carry information about  confinement as well as about $SU(2)_{CS}$  and $SU(4)$ symmetries.

Given success of \emph{CRMT} for the lowest-lying modes of the Dirac operator
it is  natural to expect that distribution law of the higher-lying
modes should be different and should reflect confinement physics.
This motivates our study of the distribution of the lowest-lying and
higher-lying modes of the Dirac operator and their comparison.

\section{Lattice setup}\label{sec-Lattice-Setup}

We compute 200 lowest eigenvalues of the overlap Dirac operator Ref. \cite{NEUBERGER1998141,NEUBERGER1998353} defined as

\begin{equation}
D_{ov} (m) = \left( \rho + \frac{m}{2} \right)  + \left( \rho - \frac{m}{2} \right) \gamma^5 sign[H(-\rho)],
\label{eq-overlap}
\end{equation}

\noindent
where $H(-\rho) = \gamma^5 D(-\rho)$ and $D(-\rho)$ is the Wilson-Dirac operator; $m = 0.015$ is the valence quarks mass and $\rho = 1.6$ is a simulation parameter. 
We use 100 gauge field configurations in the zero topological sector generated by JLQCD collaboration with $N_f = 2$ dynamical overlap fermions on a $L^3 \times L_t  = 16^3 \times 32$ lattice
with $\beta = 2.30$ and lattice spacing $a \sim 0.12fm$. The pion mass is $m_{\pi} =
289(2)MeV$, see Ref. \cite{PhysRevD.78.014508,doi:10.1093/ptep/pts006,PhysRevLett.101.202004}. 
Precisely the same gauge configurations  have
been used in truncation studies \cite{PhysRevD.89.077502,PhysRevD.91.034505,PhysRevD.91.114512,
PhysRevD.92.099902}.

We  point out that with this lattice setup we are not in the $\epsilon$-regime, because $m_{\pi}L \simeq 3$. 
Therefore it is not a priori obvious that \emph{CRMT} will work also in our case.

To obtain the eigenvalues $\lambda_{ov} (m)$ of $D_{ov}(m)$ we first calculate the sign function $sign[H]$.
We use the Chebyshev polynomials to approximate $sign[H]$ with an accuracy
of $ \epsilon = 10^{-18} $, and then  compute $ 200 $ eigenvalues of $ D_{ov} (m) $. 
For our analysis we consider only the eigenvalues with $Im(\lambda_{ov} (m)) \geq 0$ because  we know from the Ginsparg-Wilson equation $\lbrace \gamma^5,D_{ov} (0) \rbrace = \frac{1}{\rho} D_{ov}(0)\gamma^5 D_{ov}(0)$ 
and the $\gamma^5$-hermiticity, i.e. $D_{ov}^{\dagger}(0) = \gamma^5 D_{ov}(0) \gamma^5$,  that the eigenvalues come in pairs $(\lambda_{ov}(m),\lambda_{ov}^{*}(m))$. 
Therefore the eigenvalues below the real axis bring the same information as the eigenvalues above the real axis.

The eigenvalues of the overlap Dirac operator lie on a circle with radius $R = \rho - \frac{m}{2}$, see Fig. 1. 
Therefore in order to recover  chiral symmetry for the massless Dirac operator in continuum theory we need to project our eigenvalues on the imaginary axis. 
In principle there is not an unique way to do this. 
Hence we use three different projections that define the eigenvalues
in continuum limit. All these three definitions are illustrated on Fig. 1.
We will study sensitivity of our results on choice of  projection definition.

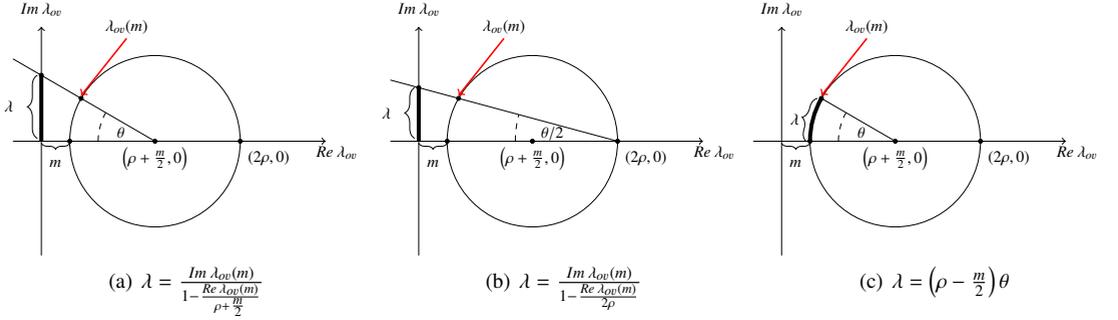
\begin{figure}[htbp]%
\centering
\subfigure[$ \lambda = \frac{Im\:\lambda_{ov}(m)}{1 - \frac{Re\:\lambda_{ov}(m)}{\rho + \frac{m}{2}}} $]{%
\label{fig-PsCenterGraph}%
\scalebox{0.75}{\begin{tikzpicture}
\draw[->] (-0.5,2.0) -- (5.0,2.0);
\draw[->] (0,0) -- (0,4.0);
\node at (5.2,1.8) {{\footnotesize $Re\:\lambda_{ov}$}};
\node at (0,4.3) {{\footnotesize $Im\:\lambda_{ov}$}};
\draw (2.0,2.0) circle (1.5);
\draw[-] (2.0,2.0) -- (0.7,2.75) -- (0,3.1538) -- (-0.5,3.4423);
\node[mark size=1pt, color=black] at (0.7,2.75){\pgfuseplotmark{*}};

\node[mark size=1pt, color=black] at (0,3.1538){\pgfuseplotmark{*}};
\node[mark size=1pt, color=black] at (2.0,2.0) {\pgfuseplotmark{*}};\node at (2.0,1.7) {{\footnotesize $\left( \rho + \frac{m}{2},0 \right)  $}};
\node[mark size=1pt, color=black] at (3.5,2.0) {\pgfuseplotmark{*}};\node at (4.0,1.7) {{\footnotesize $\left(2\rho,0\right) $}};
\node[mark size=1pt, color=black] at (0.5,2.0) {\pgfuseplotmark{*}};

\node at (1.5,4.0) {{\footnotesize $\lambda_{ov} (m)$}};


\draw[->,color=red,thick=1.6pt] (1.5,3.8) -- (0.7,2.80);


\draw[black,line width=2] (0.0,2.0) -- (0.0,3.1538);

\draw[black,dashed] (1.0,2.0) arc (180:150:1.0);\node at (1.4,2.15){{\footnotesize $\theta$}};

\draw [decorate,decoration={brace,amplitude=5pt},xshift=-2pt,yshift=0pt]
(0.0,2.05) -- (0.0,3.1538) node [black,midway,xshift=-0.5cm]{\footnotesize $\lambda$};

\draw [decorate,decoration={brace,amplitude=3pt,mirror},xshift=0pt,yshift=-1pt]
(0.0,2.0) -- (0.5,2.0) node [black,midway,yshift=-0.35cm]{\footnotesize $m$};

\end{tikzpicture}}
}%
\subfigure[$ \lambda = \frac{Im\:\lambda_{ov}(m)}{1 - \frac{Re\:\lambda_{ov}(m)}{2\rho}}$]{%
\label{fig-Ps2rhoGraph}%
\scalebox{0.75}{\begin{tikzpicture}
\draw[->] (-0.5,2.0) -- (5.0,2.0);
\draw[->] (0,0) -- (0,4.0);
\node at (5.2,1.8) {{\footnotesize $Re\:\lambda_{ov}$}};
\node at (0,4.3) {{\footnotesize $Im\:\lambda_{ov}$}};
\draw (2.0,2.0) circle (1.5);
\draw[-] (3.5,2.0) -- (0.7,2.75) -- (0,2.9375) -- (-0.5,3.0714);
\node[mark size=1pt, color=black] at (0.7,2.75){\pgfuseplotmark{*}};

\node[mark size=1pt, color=black] at (0,2.9375){\pgfuseplotmark{*}};
\node[mark size=1pt, color=black] at (2.0,2.0) {\pgfuseplotmark{*}};\node at (2.0,1.7) {{\footnotesize $\left( \rho + \frac{m}{2},0 \right)  $}};
\node[mark size=1pt, color=black] at (3.5,2.0) {\pgfuseplotmark{*}};\node at (4.0,1.7) {{\footnotesize $\left(2\rho,0\right) $}};
\node[mark size=1pt, color=black] at (0.5,2.0) {\pgfuseplotmark{*}};

\node at (1.5,4.0) {{\footnotesize $\lambda_{ov} (m)$}};

\draw[->,color=red,thick=1.6pt] (1.5,3.8) -- (0.7,2.80);


\draw[black,line width=2] (0.0,2.0) -- (0.0,2.9375);

\draw[black,dashed] (1.7,2.0) arc (180:165:1.8);\node at (2.35,2.13){{\footnotesize $\theta /2$}};

\draw [decorate,decoration={brace,amplitude=5pt},xshift=-2pt,yshift=0pt]
(0.0,2.05) -- (0.0,2.9375) node [black,midway,xshift=-0.5cm]{\footnotesize $\lambda$};

\draw [decorate,decoration={brace,amplitude=3pt,mirror},xshift=0pt,yshift=-1pt]
(0.0,2.0) -- (0.5,2.0) node [black,midway,yshift=-0.35cm]{\footnotesize $m$};
\end{tikzpicture}}
}%
\subfigure[$\lambda = \left( \rho - \frac{m}{2}\right)\theta $]{%
\label{fig-PsThetaGraph}%
\scalebox{0.75}{\begin{tikzpicture}
\draw[->] (-0.5,2.0) -- (5.0,2.0);
\draw[->] (0,0) -- (0,4.0);
\node at (5.2,1.8) {{\footnotesize $Re\:\lambda_{ov}$}};
\node at (0,4.3) {{\footnotesize $Im\:\lambda_{ov}$}};
\draw (2.0,2.0) circle (1.5);
\draw[-] (2.0,2.0) -- (0.7,2.75);
\node[mark size=1pt, color=black] at (0.7,2.75){\pgfuseplotmark{*}};

\node[mark size=1pt, color=black] at (2.0,2.0) {\pgfuseplotmark{*}};\node at (2.0,1.7) {{\footnotesize $\left( \rho + \frac{m}{2},0 \right)  $}};
\node[mark size=1pt, color=black] at (3.5,2.0) {\pgfuseplotmark{*}};\node at (4.0,1.7) {{\footnotesize $\left(2\rho,0\right) $}};
\node[mark size=1pt, color=black] at (0.5,2.0) {\pgfuseplotmark{*}};

\node at (1.5,4.0) {{\footnotesize $\lambda_{ov} (m)$}};

\draw[->,color=red,thick=1.6pt] (1.5,3.8) -- (0.7,2.80);


\draw[black,line width=2] (0.5,2.0) arc (180:150:1.5);

\draw[black,dashed] (1.0,2.0) arc (180:150:1.0);\node at (1.4,2.15){{\footnotesize $\theta$}};

\draw [decorate,decoration={brace,amplitude=5pt},xshift=-2pt,yshift=0pt]
(0.5,2.0) -- (0.7,2.75) node [black,midway,xshift=-0.3cm]{\footnotesize $\lambda$};

\draw [decorate,decoration={brace,amplitude=3pt,mirror},xshift=0pt,yshift=-1pt]
(0.0,2.0) -- (0.5,2.0) node [black,midway,yshift=-0.35cm]{\footnotesize $m$};
\end{tikzpicture}}
}%
\label{fig-ProjDefinition}
\caption{Different definitions of $\lambda$  using the eigenvalues of the overlap Dirac operator, $\lambda_{ov} (m)$.
The angle $\theta$ is defined as $\theta = arctg\left( \frac{Im\: \lambda_{ov}(m)}{\left( \rho + \frac{m}{2}\right) - Re\: \lambda_{ov}(m)}\right) $.}
\end{figure}

\section{Lowest eigenvalues}\label{sec-Lowest-eigenvalues}

First we study the lowest eigenvalues of the Dirac operator. 
We are not in the $\epsilon$-regime so we don't expect a priori a full agreement with \emph{CRMT}. Consequently we want to check whether and to what extent
the lowest eigenvalues are described by \emph{CRMT}.

An important prediction of \emph{CRMT} is the distribution of the lowest eigenvalues in the thermodynamic limit. 
It means in the limit when the four dimensional volume of our lattice $V\rightarrow \infty$ and the quantity $V\Sigma m_{\pi}$ is fixed. 

Defining the variables $\zeta_k  = V\Sigma \lambda_k$, where, in our case, $\lambda_k$ is the $k$-th lowest projected eigenvalue of the overlap Dirac operator, 
then we can get (see Ref. \cite{PhysRevD.63.045012}) the distribution $p_k (\zeta_k)$ of each $\zeta_k$. 

Since we don't know the parameter $\Sigma$ 
 we study the ratios $\langle \lambda_k \rangle/\langle \lambda_j \rangle$,
where $\langle \lambda_i \rangle$ is the average over all gauge configurations for the $i$th projected eigenvalue. 
Indeed using that $\langle \zeta_k \rangle = V\Sigma \langle\lambda_k \rangle$, we can compare our ratios with the predictions of \emph{CRMT}. 
We show the data for $1 \leq j < k \leq 4$ in Table 1 and in the last column there are the values predicted by \emph{CRMT}. 

\begin{table}[thb]


  \caption{Ratio $\langle \lambda_k \rangle / \langle \lambda_j \rangle$ for $1\leq j \leq k \leq 4$ and the same values computed with the \emph{CRMT}. 
  We denote with $\sigma$ the error. 
  In this case we have used $\lambda$ defined as in Fig. \ref{fig-Ps2rhoGraph}.}
  \label{tab-EigRatio}
  \begin{tabular}{llll}\toprule
  $ k/j $ & $\langle \lambda_k \rangle / \langle \lambda_j \rangle$ & $\sigma$  & \emph{CRMT} \\\midrule
  2/1 & 2.72 &	0.19 & 2.70 \\
  3/1 & 4.35 &	0.28 & 4.46 \\
  3/2 & 1.60 &	0.06 & 1.65 \\
  4/1 & 5.92 &	0.38 & 6.22 \\
  4/2 & 2.17 &	0.08 & 2.30 \\
  4/3 & 1.36 &	0.03 & 1.40 \\\bottomrule
  \end{tabular}
\end{table}

\begin{figure}[thb]
\centering
\includegraphics[scale=0.4]{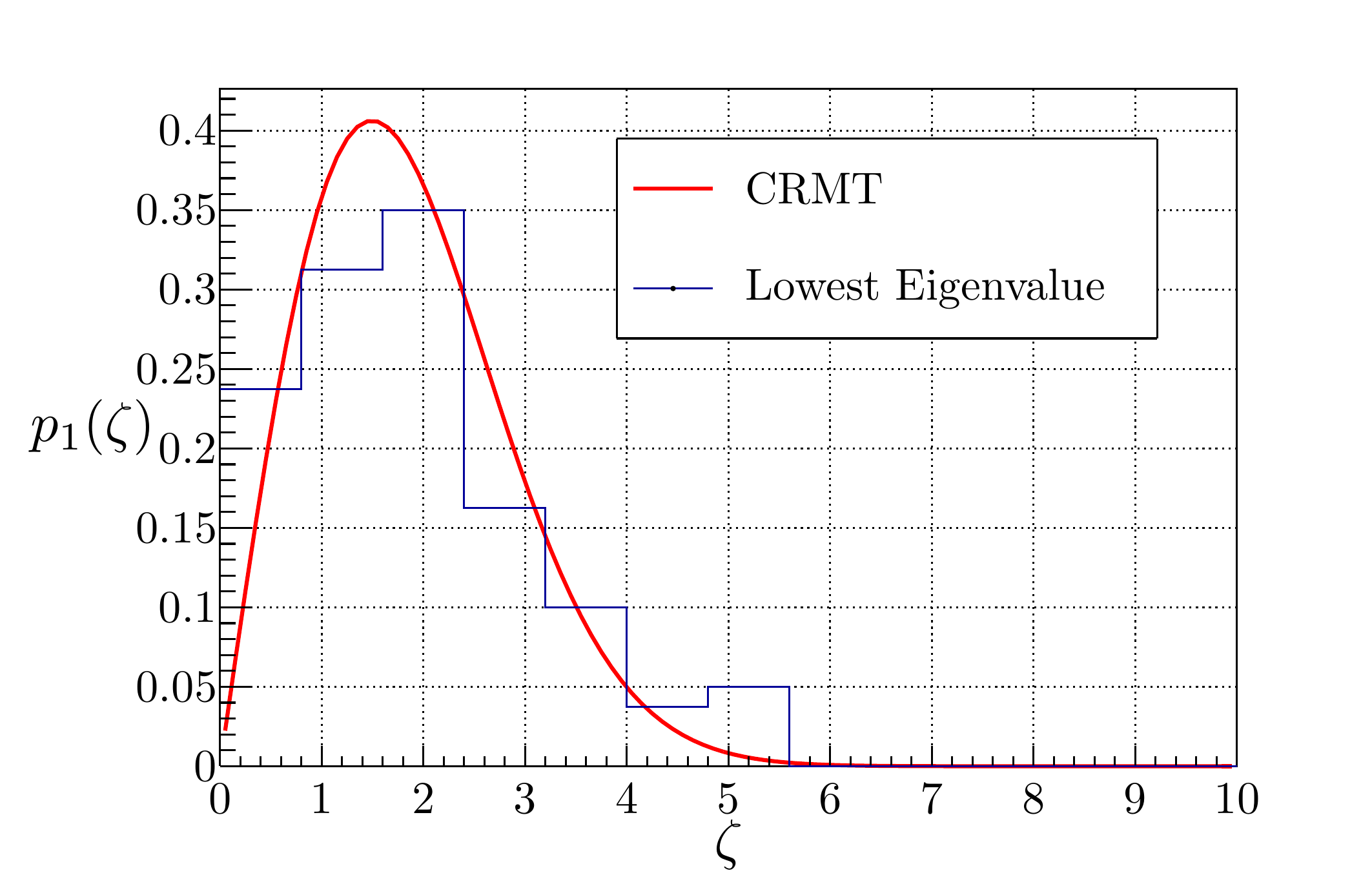}
\caption{Distribution of the lowest eigenvalue. In this case $\zeta = V\Sigma \lambda$.}
\label{fig-Distr1stEig}

\end{figure}

We can see that the ratios for the first $3$ projected eigenvalues are in good agreement with the \emph{CRMT}. 
The ratios involving the $4$-th projected eigenvalues have a larger discrepancy.

From the theoretical values of $\langle \zeta_k \rangle$ and the observed values of $\langle \lambda_k \rangle$ we can get the parameter $\Sigma$. 
We find for our system that $\Sigma = (232.2 \pm 0.9 MeV)^3$. 
We use this parameter to compare the distribution of the first lowest projected eigenvalue with the theoretical distribution given by \emph{CRMT}, 
as we report in Fig. \ref{fig-Distr1stEig}.

We can conclude this section observing that, even if we are not in the $\epsilon$-regime, for very low eigenvalues in our system the predictions of \emph{CRMT} still work.

\section{Nearest neighbor spacing distribution}\label{sec-NNS}

Another important prediction of \emph{CRMT} is the \emph{nearest neighbor spacing distribution} (or \emph{NNS} distribution). 
This is the distribution of the variable 

$$s_n  = \xi_{n+1} - \xi_{n},$$ 

\noindent
where 

$$\xi_n = \xi (\lambda_n) = \int_{0}^{\lambda_n} R(\lambda)d\lambda$$ 

\noindent
and $R(\lambda)d\lambda$ is the probability to find an eigenvalue of the Dirac operator in the interval $(\lambda,\lambda + d\lambda)$. 
$n$ indicates the number of the lowest projected eigenvalue, 
supposing we have ordered the projected eigenvalues such that $\lambda_1 \leq \lambda_2 \leq ... \leq \lambda_n$. 
In principle we don't know how $R(\lambda)$ is made and the procedure to map the set of variables $\lbrace \lambda_1, ...,\lambda_n \rbrace$ 
into the set $\lbrace \xi_1,...,\xi_n \rbrace$ is called \emph{unfolding} and it is described in \cite{Guhr1998189}. 
This procedure is based on the introduction of the following variable 

$$
\eta (\lambda_n) = \int_{0}^{\lambda_n} \rho (\lambda)d\lambda = \frac{1}{N} \langle \sum_k \theta (\lambda_n - \lambda_k) \rangle = 
\frac{1}{N}\frac{1}{M}\sum_{i=1}^M \sum_k \theta (\lambda_n - \lambda_k^i),
$$

\noindent
where $\rho (\lambda) =  \frac{1}{N} \langle \sum_k \delta (\lambda - \lambda_k) \rangle$ is the spectral density of the Dirac operator averaged over all gauge field configurations, 
$\lambda_k^i$ denotes the $k$th lowest projected eigenvalue of the Dirac operator computed using the $i$th gauge configuration. 
$M=100$ is the number of gauge configurations that we are taking into account and $N$ is total number of the eigenvalues of the Dirac operator.
Now $\eta (\lambda)$ can be decomposed in a fluctuating part $\eta_{fl}(\lambda)$ and in a smooth part $\xi (\lambda)$, namely $\eta (\lambda) = \xi (\lambda) + \eta_{fl}(\lambda)$. 
The smooth part $\xi (\lambda)$ can be obtained by a polynomial fit of $\eta (\lambda)$.

For different values of the Dyson index $\beta$ we have different shapes for the \emph{NNS} distribution. 

We use this distribution to study the lowest and higher  eigenvalues of the overlap Dirac operator.  
The lowest eigenvalues contain the information about the $SU(2)_L \otimes SU(2)_R$ and $U(1)_A$ breakings. The \emph{NNS} distribution calculated with
10 lowest Dirac eigenmodes is shown in the left panel of Fig. 3. We
see that the distribution is perfectly described by the Gaussian Unitary Ensemble, in agreement with \emph{CRMT}.

The right panel of Fig. 3 shows  \emph{NNS} distribution
obtained with eigenmodes in the interval 81 - 100. It is clear from
results of refs. \cite{PhysRevD.89.077502,PhysRevD.91.034505,PhysRevD.91.114512,
PhysRevD.92.099902} that this part of the Dirac spectrum is not sensitive
to \emph{SBCS} and to breaking of $U(1)_A$, but reflect  physics of
confinement and of   $SU(2)_{CS}$ and $SU(4)$ symmetries. Nevertheless,
distribution of these eigenmodes of the Dirac operator is described
by the same Wigner distribution (\emph{GUE}) as of the lowest ten modes, which is unexpected.
 
Finally, in Fig. 4 we show \emph{NNS} distributions of the lowest 100 modes
calculated with three different definitions of projected eigenvalue,
compare with Fig. 1.
It is clear that results for distribution is not sensitive to definition of
projected eigenvalue.

\begin{figure}[!thb]
 \centering
 \subfigure[Range eigenvalues: 1 - 10]{
 \label{fig-1-10}
 \includegraphics[scale=0.32]{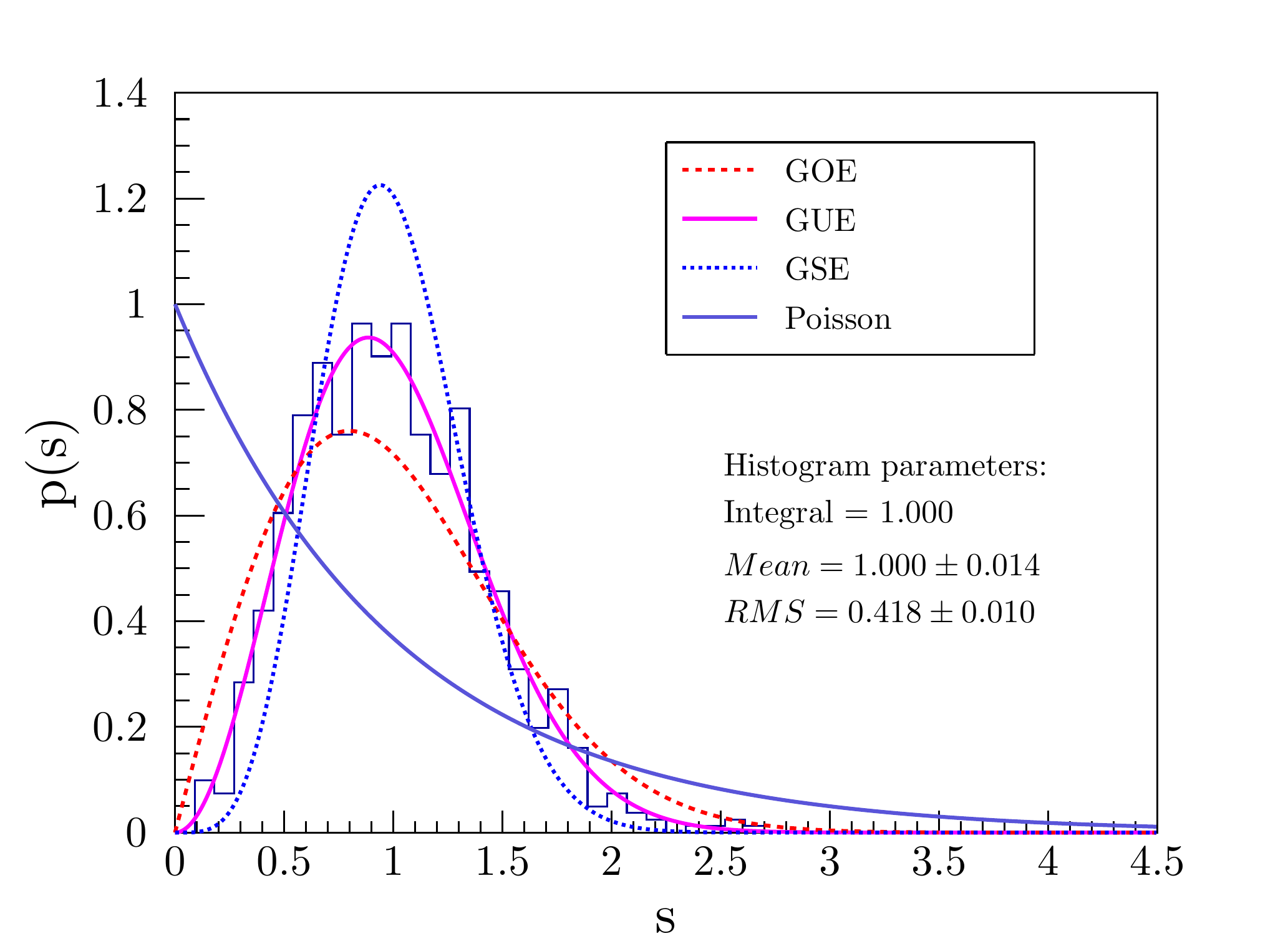}
 }
  \subfigure[Range eigenvalues: 81 - 100]{
  \label{fig-81-100}
  \includegraphics[scale=0.32]{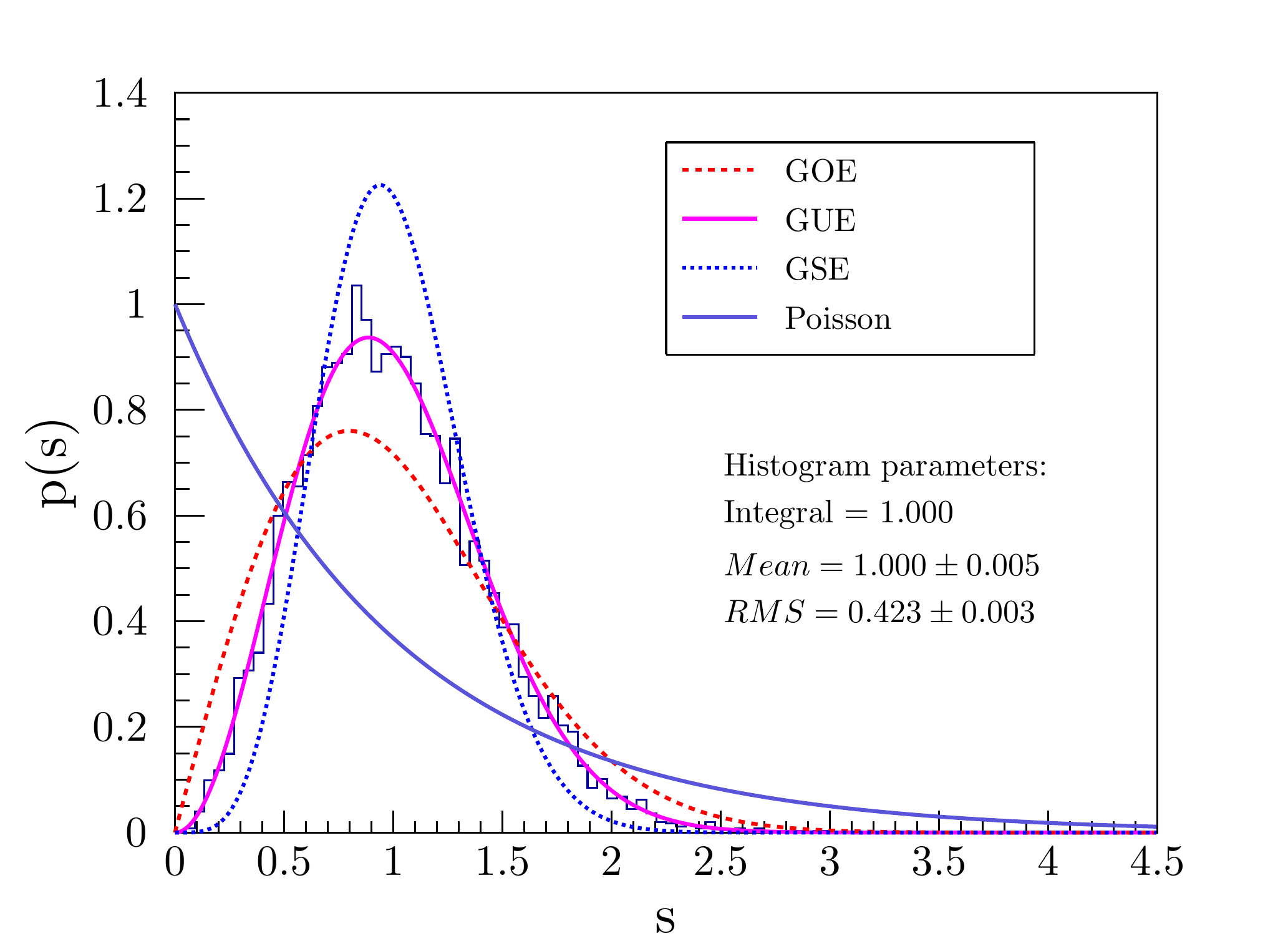}
 }
 \caption{\emph{NNS} distribution for the first $10$ lowest projected eigenvalues of the overlap Dirac operator on the left, and the higher eigenvalues on the right. 
 We have used $\lambda$ defined as in Fig. 1(b).}
 \label{fig-NNS_distribution}
\end{figure}

\begin{figure}[!thb]
 \centering
 \subfigure[$ \lambda = \frac{Im\:\lambda_{ov}(m)}{1 - \frac{Re\:\lambda_{ov}(m)}{\rho + \frac{m}{2}}} $]{
 \label{fig-PsCenter}
 \includegraphics[scale=0.32]{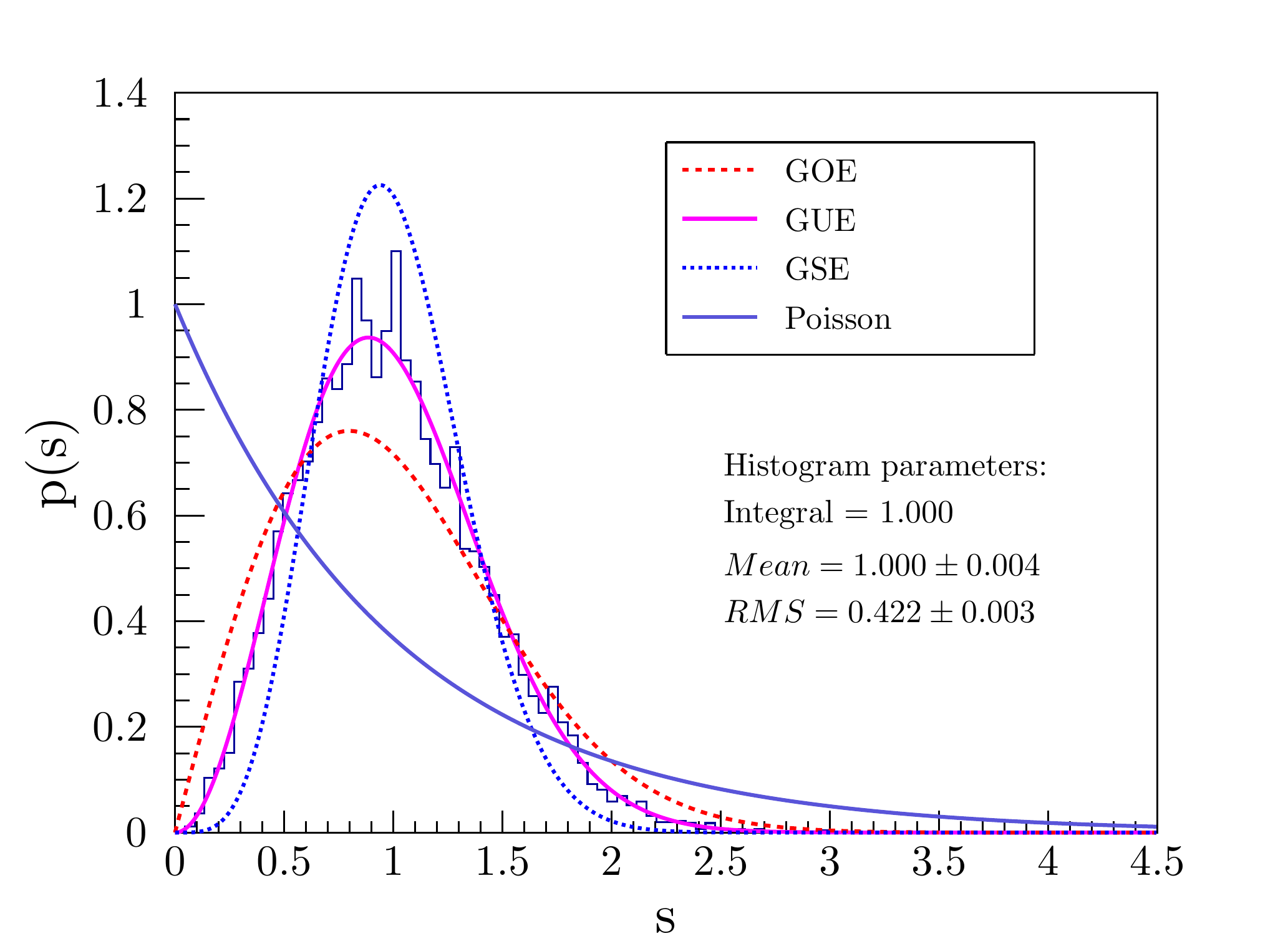}
 }
 \subfigure[$ \lambda = \frac{Im\:\lambda_{ov}(m)}{1 - \frac{Re\:\lambda_{ov}(m)}{2\rho}}$]{
 \label{fig-Ps2rho}
 \includegraphics[scale=0.32]{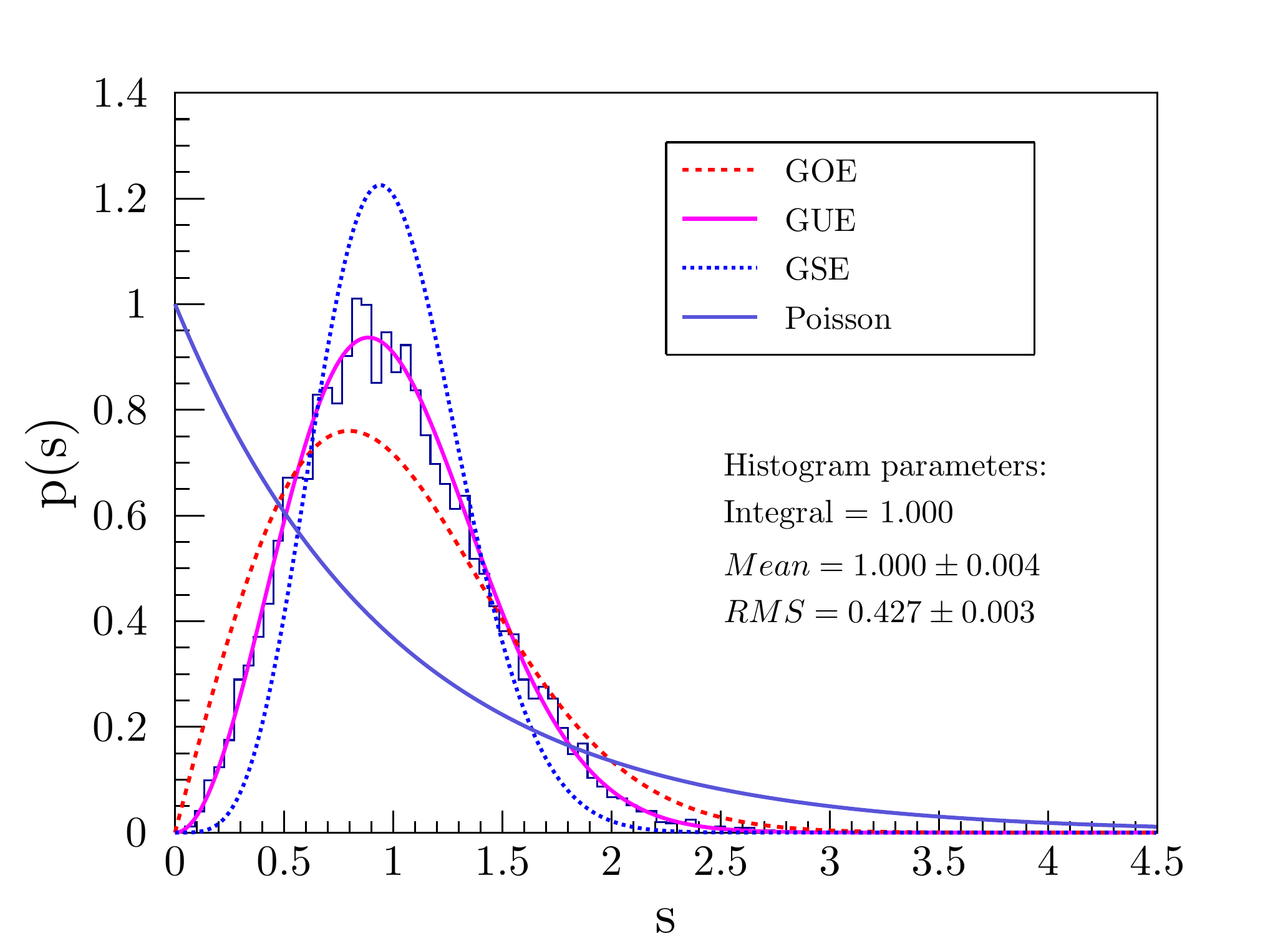}
 }
 \subfigure[$\lambda = \left( \rho - \frac{m}{2}\right)\theta $]{
 \label{fig-PsTheta}
 \includegraphics[scale=0.32]{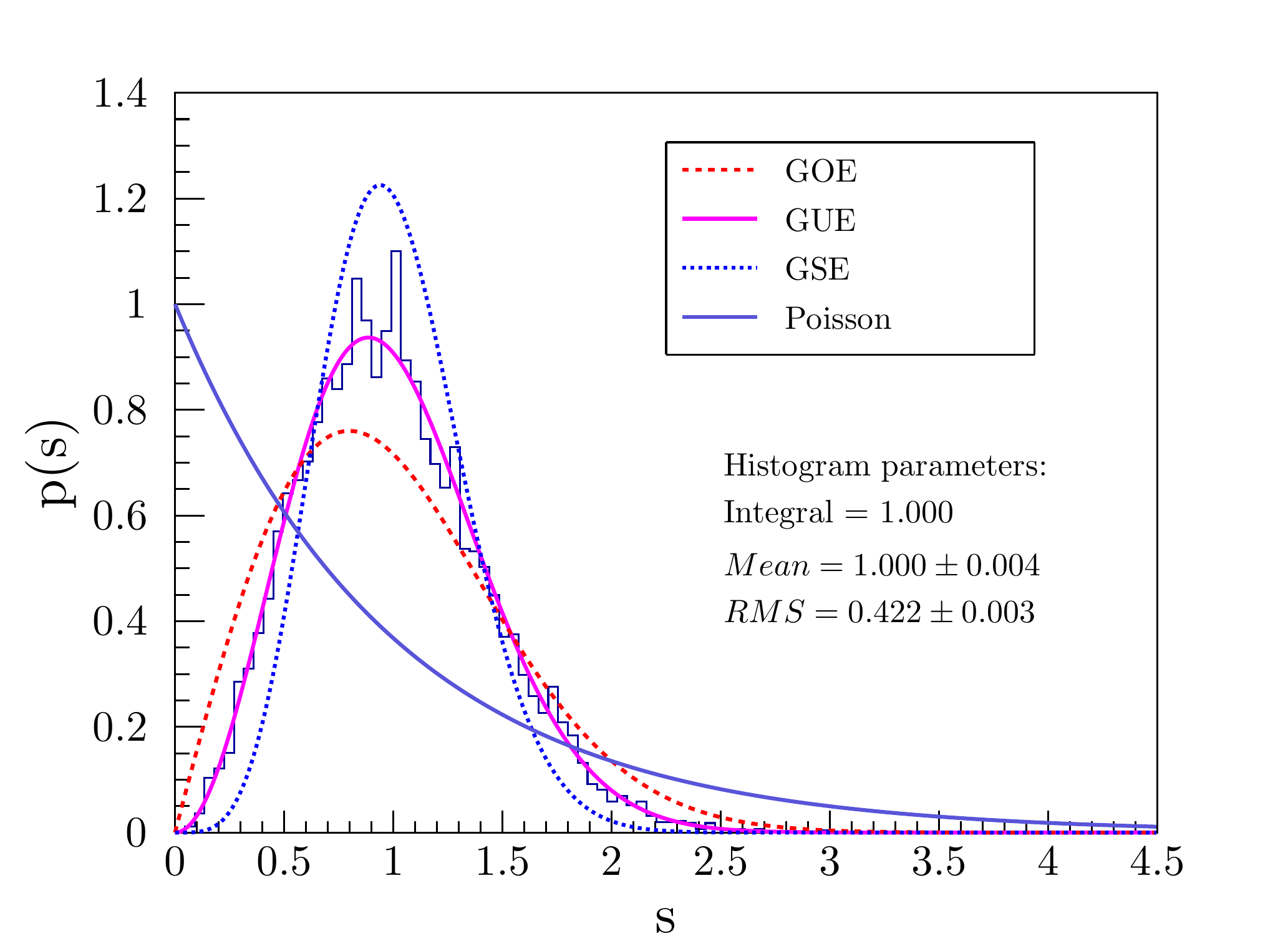}
 }
  \caption{\emph{NNS} distribution for the lowest $100$ eigenvalues of the Dirac operator, for different definitions of $\lambda$.}
 \label{fig-NNS_distribution_different_projections}
\end{figure} 
 
\section{Conclusions}\label{sec-Conclusions}

In this work we have analyzed distributions  of
the lowest and the higher-lying eigenvalues of the overlap Dirac operator.  
We have seen that the lowest eigenvalues are well described by \emph{CRMT}
in agreement with previous studies.

We have also studied the nearest neighbor spacing distribution for  the higher eigenvalues that are  affected neither by $U(1)_A$ breaking nor by spontaneous
breaking of chiral symmetry. These modes are sensitive to confinement
physics and to related $SU(2)_{CS}$ and $SU(4)$ symmetries.
We have found that they follow the Wigner distribution as the near-zero modes. 
This observation means that the Wigner distribution seen  both for
the near-zero and higher-lying modes,
while consistent with spontaneous breaking of chiral symmetry,
is not a consequence of spontaneous breaking of chiral symmetry in
QCD but has some more general origin in QCD in confinement regime.
In other words a randomness that we observe both for the near-zero modes
and for the higher-lying modes has not yet known origin in QCD.

 \medskip
 \noindent
{\bf Acknowledgments}

We thank C.B. Lang for numerous discussions.
This work is supported by the Austrian Science Fund FWF through grants  
DK W1203-N16 and P26627-N27.
\bibliography{lattice2017}

\end{document}